\documentclass[conference]{IEEEtran}
\IEEEoverridecommandlockouts
\usepackage{cite}
\usepackage{amsmath,amssymb,amsfonts}
\usepackage{algorithmic}
\usepackage{graphicx}
\usepackage{textcomp}
\usepackage{physics}
\usepackage{xcolor}
\usepackage{hyperref}
\def\BibTeX{{\rm B\kern-.05em{\sc i\kern-.025em b}\kern-.08em
    T\kern-.1667em\lower.7ex\hbox{E}\kern-.125emX}}

\newcommand{\vect}[1]{\boldsymbol{\mathbf{#1}}}

\begin{document}

\title{Adaptive Quantum Generative Training using an Unbounded Loss Function\\
\thanks{This work was funded by grants from the US Department of Energy, Office of Science, National Quantum Information Science Research Centers, Co-Design Center for Quantum Advantage under contract number DE-SC0012704.}
}

\makeatletter
\newcommand{\linebreakand}{%
  \end{@IEEEauthorhalign}
  \hfill\mbox{}\par
  \mbox{}\hfill\begin{@IEEEauthorhalign}
}
\makeatother

\author{\IEEEauthorblockN{Kyle Sherbert}
\IEEEauthorblockA{\textit{Department of Physics} \\
\textit{Virginia Tech}\\
Blacksburg, VA \\
kyle.sherbert@vt.edu}
\and
\IEEEauthorblockN{Jim Furches$^*$}
\IEEEauthorblockA{\textit{Department of Physics} \\
\textit{Virginia Tech}\\
Blacksburg, VA \\
jfurches@vt.edu}
\and
\IEEEauthorblockN{Karunya Shirali$^*$}
\IEEEauthorblockA{\textit{Department of Physics} \\
\textit{Virginia Tech}\\
Blacksburg, VA \\
karunyashirali@vt.edu}
\linebreakand
\IEEEauthorblockN{Sophia E. Economou}
\IEEEauthorblockA{\textit{Department of Physics} \\
\textit{Virginia Tech}\\
Blacksburg, VA \\
economou@vt.edu}
\and
\IEEEauthorblockN{Carlos Ortiz Marrero}
\IEEEauthorblockA{\textit{AI \& Data Analytics Division} \\
\textit{Pacific Northwest National Laboratory}\\
Seattle, WA \\
carlos.ortizmarrero@pnnl.gov}
}

\maketitle

\def\thefootnote{*}\footnotetext{These authors contributed equally to this work}

\begin{abstract}
We propose a generative quantum learning algorithm, Rényi-ADAPT, using the Adaptive Derivative-Assembled Problem Tailored ansatz (ADAPT) framework in which the loss function to be minimized is the maximal quantum Rényi divergence of order two, an unbounded function that mitigates barren plateaus which inhibit training variational circuits. We benchmark this method against other state-of-the-art adaptive algorithms by learning random two-local thermal states. We perform numerical experiments on systems of up to 12 qubits, comparing our method to learning algorithms that use linear objective functions, and show that Rényi-ADAPT is capable of constructing shallow quantum circuits competitive with existing methods, while the gradients remain favorable resulting from the maximal Rényi divergence loss function.
\end{abstract}


\section{Introduction}
Many promising quantum machine learning algorithms almost exclusively use a linear bounded operator as their objective functions during training \cite{Peruzzo2014, farhi2014quantum, havlivcek2019supervised, schuld2020circuit, ceroni2022generating, Feniou_2023}. These loss functions are typically estimated by measuring the expectation values of Hermitian operators directly during training, computing their gradients, and updating the parameters. Despite optimism, training using linear objective functions has proved computationally difficult due to a concentration of measure phenomenon known as a \emph{barren plateau} that leads to exponentially small gradients as we scale the system size of our problem \cite{larocca2024review}. These results paint a bleak picture for the future of quantum machine learning, and finding a scalable approach to train generic parametrized models has become a central problem in the field.  Existing approaches to overcome barren plateaus are based on empirical evidence~\cite{grant2019initialization, skolik2021layerwise} or are restricted to a specific architecture~\cite{pesah2020absence, cerezo2020cost, sharma2020trainability} but none provide a generic way to train quantum models that is guaranteed to avoid these no-go results. Recent work by Kieferov{\'a} et al. \cite{kieferova2021quantum} suggests the use of a Quantum Divergence, an {\it unbounded}, nonlinear loss function with simple gradients, in place of a linear objective function, when performing generative training to address the barren plateau problem. The standard arguments for barren plateau theorems do not apply to divergences because these experience a logarithmic divergence when the two states are nearly orthogonal. This causes the gradients of the divergence between nearly orthogonal quantum states to be large, and thereby provides a workaround for all known barren plateau results. 

In our work, we extend Kieferov{\'a} et al. \cite{kieferova2021quantum} method and propose an adaptive training algorithm to optimize the maximal quantum Rényi divergence between two quantum states, that is, the output of a quantum circuit and some quantum data state. We provide evidence showing how our adaptive algorithm performs compared to other generative training algorithms. In particular, we explore the gradient evolution of each method and argue that our {\it Rényi-ADAPT} has favorable gradients at scale, when compared to methods such as Overlap-ADAPT \cite{Feniou_2023} and ADAPT-VQE-Gibbs\cite{warren2022adaptive}. 

\section{ADAPT-VQE algorithm}\label{adapt-sec}
Variational quantum eigensolvers (VQEs)~\cite{Cerezo2021,Tilly2022,Bharti2022} employ parametrized quantum circuits to divide the computational task of optimizing an objective function $f(\theta_1, \theta_2, \dots \theta_n)$ (most often the ground state energy of a physical system, estimated as the expectation value of the system Hamiltonian $\langle H \rangle$) between quantum and classical processors, in which the quantum processor performs measurements of the objective function and the classical processor updates the variational parameters $\{\theta_i\}$ at each iteration. The choice of parametrized circuit (the ``ansatz'') is critical to the performance of the algorithm, and the limited coherence times available on current quantum hardware prompt the use of shallow circuits. In general, a good ansatz needs to be sufficiently expressive to represent the solution accurately, but also have a low circuit depth. 

The Adaptive Derivative-Assembled Problem Tailored ansatz (ADAPT)-VQE algorithm~\cite{Grimsley2019a} employs an adaptive strategy to construct the variational ansatz dynamically using operators in a predefined operator pool, starting from an initial state $|\psi_{\rm ref}\rangle$. Defining the variational parameters $\vect{\theta}^{(k)} = (\theta_1,\dots,\theta_k)$ and the operator pool $\mathcal{A} = \{ A^{(1)}, A^{(2)}, \dots A^{(N)}\}$, the ansatz in iteration $k+1$ of the algorithm may be written as
\begin{equation}\label{adapt-vqe-eqn}
    |\psi_{k+1}(\vect{\theta}^{(k+1)})\rangle = e^{-i\theta_{k+1 }A_{k+1}}|\psi_{k}(\vect{\theta}^{(k)})\rangle.
\end{equation}
In (\ref{adapt-vqe-eqn}), the ansatz at iteration $k$ is grown by appending operator $A_{k+1}$ with coefficient $\theta_{k+1}$; the operator is chosen by measuring the energy gradients $\left| \left. \partial \langle H \rangle / \partial \theta_{k+1}\right|_{\theta_{k+1}=0} \right|$ for each operator in the pool and selecting the one with the largest gradient. For this step, it can be shown that 
\begin{equation}\label{adapt-gradient}
    \left| \left. \partial \langle H \rangle / \partial \theta_{k+1}\right|_{\theta_{k+1}=0} \right| =  \left| \langle \psi_{k}(\vect{\theta}^{(k)}) | \left[ A_{k+1}, H \right] | \psi_{k}(\vect{\theta}^{(k)}) \rangle \right|,
\end{equation}
where the right hand side of (\ref{adapt-gradient}) can be efficiently measured on a quantum processor. The pool operator gradient-measurement step is followed by a convergence check: if the pool operator gradient norm is smaller than a threshold $\varepsilon$, the calculation is terminated; if not, the ansatz is grown as in (\ref{adapt-vqe-eqn}). The ansatz-growing step is followed by a VQE optimization of all variational parameters, where 
the parameters $(\theta_1,\dots,\theta_k)$ take their previously optimized values, and the newly added parameter is initialized as $0$.

In this work, we are interested in training mixed states, so we extend ADAPT following the strategy in \cite{warren2022adaptive},
    in which we use \eqref{adapt-vqe-eqn} to prepare a pure state on a set of {\em visible} and {\em hidden} (ancilla) qubits.
The trial mixed state is taken as the state on the visible qubits when the hidden qubits are traced out:
\begin{equation}\label{trace-out-hidden}
    \sigma_k(\vect{\theta}^{(k)}) = \Tr_H{\ketbra{\psi_k(\vect{\theta}^{(k)})}}
\end{equation}
We hereafter write $\sigma_k(\vect{\theta}^{(k)})$ as $\sigma(\theta)$ for notational simplicity, or even simply as $\sigma$ when context allows.

The ADAPT-VQE algorithm was shown to reach arbitrarily accurate energies and yield shallow ansätze when applied to small molecules, and inspired adaptive techniques for studying other problems, including time evolution and dynamics~\cite{Zhang2021,Gomes2021}, excited-state preparation~\cite{Yao2021}, and optimization~\cite{LZhu2020,chen2023entanglement}. In particular, Overlap-ADAPT-VQE~\cite{Feniou_2023}, a method that we use in this work, dynamically constructs an ansatz by maximizing the overlap between a trial state and the target state. It was found to produce ultracompact ansätze, thereby providing accurate reference states for subsequent ADAPT runs, and resulting in shallower circuit depths of the final ansatz compared to regular ADAPT. 

The loss function in Overlap-ADAPT-VQE~\cite{Feniou_2023}, formulated for pure states, is 
\begin{equation}\label{original-overlap-adapt-eqn}
    \mathcal{L}_O(\theta) = \langle \Psi_{\rm target} | \psi(\theta) \rangle.
\end{equation}
For mixed states $\rho$ and $\sigma$, the fidelity has the following generalization:
\begin{equation}\label{mixed-state-fidelity}
    F(\rho, \sigma) = \text{Tr} \sqrt{\sqrt{\rho} \, \sigma(\theta) \, \sqrt{\rho}}
\end{equation}
In this work, we extend Overlap-ADAPT-VQE to prepare mixed states, using the loss-function $1 - F(\rho, \sigma(\theta))^2$,
    where $\rho$ is the target state and $\sigma(\theta)$ is the trial state.
\begin{equation}\label{overlap-adapt-loss-fn}
    \mathcal{L}_O(\rho, \sigma(\theta)) = 1 - \left(\text{Tr} \sqrt{\sqrt{\rho} \, \sigma(\theta) \, \sqrt{\rho}}\right)^2. 
\end{equation}

\section{Thermal state preparation using the ADAPT-VQE-Gibbs algorithm}
An adaptive protocol for the preparation of Gibbs thermal states $\rho_G = e^{-\beta H}/Z$ was proposed in~\cite{warren2022adaptive} in which an easily measurable customized objective function $C(\sigma)$ was introduced; it replaced the Gibbs free energy $F(\sigma) = E(\sigma) - k_B TS(\sigma)$ as the objective function to minimize. Using this objective function, in combination with the ADAPT-VQE strategy of dynamically constructing ansätze from a predefined operator pool, it was shown that accurate representations of the Gibbs state could be obtained across a range of temperatures using low-depth circuits. The protocol makes use of a purifying ancilla system $A$ in addition to the data system $D$ to generate a pure state $|\psi\rangle \in \mathcal{H}_D \otimes \mathcal{H}_A$, where the Gibbs state $\rho_G$ of system $D$ is obtained by tracing out the ancillary system (the visible and hidden qubits in this work correspond to the data and ancilla qubits, respectively, in ADAPT-VQE-Gibbs). The modified objective function
\begin{equation}\label{gibbs-loss-fn}
C(\sigma(\theta)) = - \text{Tr}(\rho_G \sigma) + \frac{1}{2} \text{Tr}(\sigma^2),
\end{equation}
where $\sigma = \text{Tr}_A |\psi\rangle\langle\psi|$ is the variational trial state, and $\rho_G$ is an approximation of the target Gibbs state. Using $C(\sigma)$ as the objective function is convenient because it circumvents the need to measure the von Neumann entropy, which is hard to measure on current quantum hardware. The approximation of $\rho_G$ in (\ref{gibbs-loss-fn}) is obtained via a truncated Taylor expansion of $e^{-\beta H}/Z$.

In our calculations, we truncate the Taylor expansion of $e^{-\beta H}$ at order $m=5$, i.e., set $e^{-\beta H} \approx \sum_{n=0}^{5} \frac{1}{n!} (-\beta)^n H^n$, which was found to lead to Gibbs states of fidelities comparable to those obtained when using the exact operator produced by summing the infinite series~\cite{warren2022adaptive}.

\section{Rényi-ADAPT}

The loss function which we are interested in comparing is a generalization of the quantum relative entropy known as the quantum Rényi divergence or ``sandwiched'' Rényi relative entropy ~\cite{wilde2014strong, muller2013quantum, petz1986quasi}. The Quantum Rényi divergence inherits many of the mathematical properties of the Rényi divergence, and it reduces to the quantum relative entropy (quantum analog of the KL-divergence) and the limit where the dominant parameter of the loss function goes to 1.
We are particularly interested in using an upper bound to the quantum Rényi divergence called the maximal Rényi divergence of order 2, 
\begin{equation}\label{renyi-loss-fn-reversed}
    \widetilde{D}_{2}(\sigma(\theta)|\rho) = \log\left({\rm Tr}\left(\sigma^2 \rho^{-1}\right) \right).
\end{equation}
Here, $\rho$ is the density of the training data state and $\sigma(\theta)$ corresponds to the density of the output state of a parameterized quantum circuit, where $\theta$ are the circuit parameters. These loss functions have not been widely studied in a quantum computing context and were introduced in \cite{kieferova2021quantum} as a way to do quantum generative training. 
The main reason for using $\widetilde{D}_2(\sigma|\rho)$ as a loss function is that it upper-bounds the quantum relative entropy and its gradients are considerably simpler to measure on quantum hardware than those of the ordinary Rényi divergence and quantum relative entropy \cite{kieferova2021quantum}. Please consult Appendix \ref{gradients} for more information on the gradients.
Note that, in general, $\widetilde{D}_{2}(\rho|\sigma) \neq \widetilde{D}_{2}(\sigma|\rho)$. However, if both $\rho$ and $\sigma$ are full rank, $\widetilde{D}_{2}(\rho|\sigma) = \widetilde{D}_{2}(\sigma|\rho) = 0$ if and only if $\rho=\sigma$~\cite{renyi}. This makes $\widetilde{D}_{2}(\sigma(\theta)|\rho)$ is a reasonable loss function to consider for generative training.




As described in Section \ref{adapt-sec}, the ADAPT algorithm selects an operator from a pool with the largest energy gradient at each iteration. We can apply a similar logic, but instead use the gradient of the maximal R\'enyi divergence to select operators from a pool $\{H_1, \dots, H_k\}$ to form a trial state $\sigma(\theta)$ following the logic outlined in equations~\eqref{adapt-vqe-eqn} and~\eqref{trace-out-hidden}, which we can then optimize at each iteration. This technique is what we call Rényi-ADAPT.

\section{Experiments}

Following the notation of \cite{kieferova2021quantum}, we denote the $n$-qubit target thermal state as $\rho$, the controllable parameters as $\theta$, and the trial state as $\sigma(\theta)$. In the context of the algorithm, the trial state starts as a pure state acting on a register of $n_V$ visible qubits and $n_H$ hidden (ancilla) qubits, within a full Hilbert space of $\mathcal{H}_\sigma = \mathcal{H}_V \otimes \mathcal{H}_H$.
The final (mixed) state output of $\sigma(\theta)$ is obtained by tracing out the hidden register.
Unless otherwise noted, all experiments were carried out with $n_H = n_V$, and we use the letter $n$ to denote both.
All multivariate optimization steps were performed using the BFGS algorithm, as implemented in the \verb|Optim.jl| Julia package \cite{optim,julia}.

\subsection{Adaptive optimization}

In order to demonstrate that the adaptive optimization process works, we ran ADAPT on a single random two-local thermal state $\rho$ of $n = 3$ qubits.
We trained ans\"atze with three different loss functions (\ref{overlap-adapt-loss-fn},\ref{gibbs-loss-fn},\ref{renyi-loss-fn-reversed}) to learn $\rho$.
For the operator pool, we include all one- and two-local Paulis acting on any of the six qubits (three visible and three hidden).
There are a total of ${6 \choose 1}\cdot3^1=18$ one-local operators and ${6\choose2}\cdot3^2=135$ two-local operators, for a total of 153 operators.
In addition to the three ADAPT variants investigated, we ran VQE with a fixed-structure ansatz (``VQE'') consisting of all 153 Pauli rotations (with arbitrary ordering) for all three loss functions, to compare against previous work \cite{kieferova2021quantum}.

\begin{figure}
    \includegraphics[width=\columnwidth]{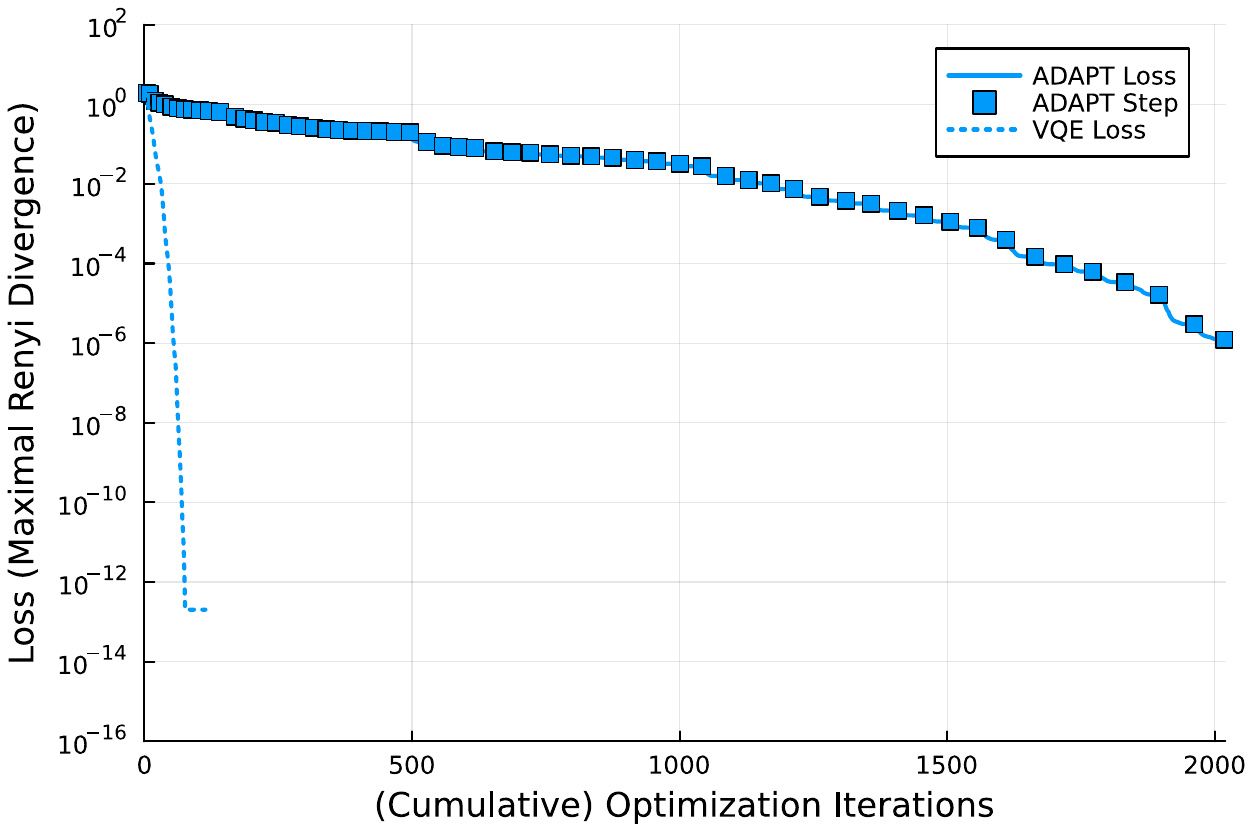}
    \includegraphics[width=\columnwidth]{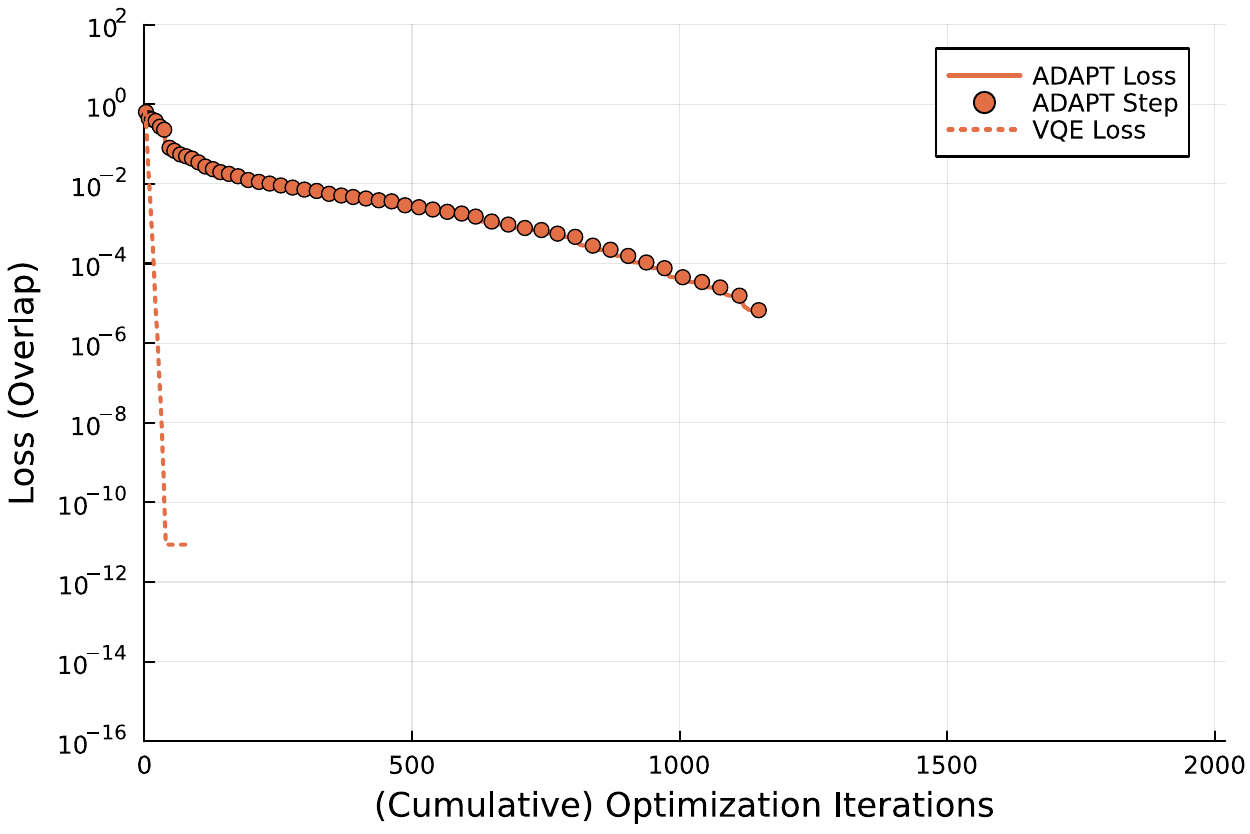}
    \includegraphics[width=\columnwidth]{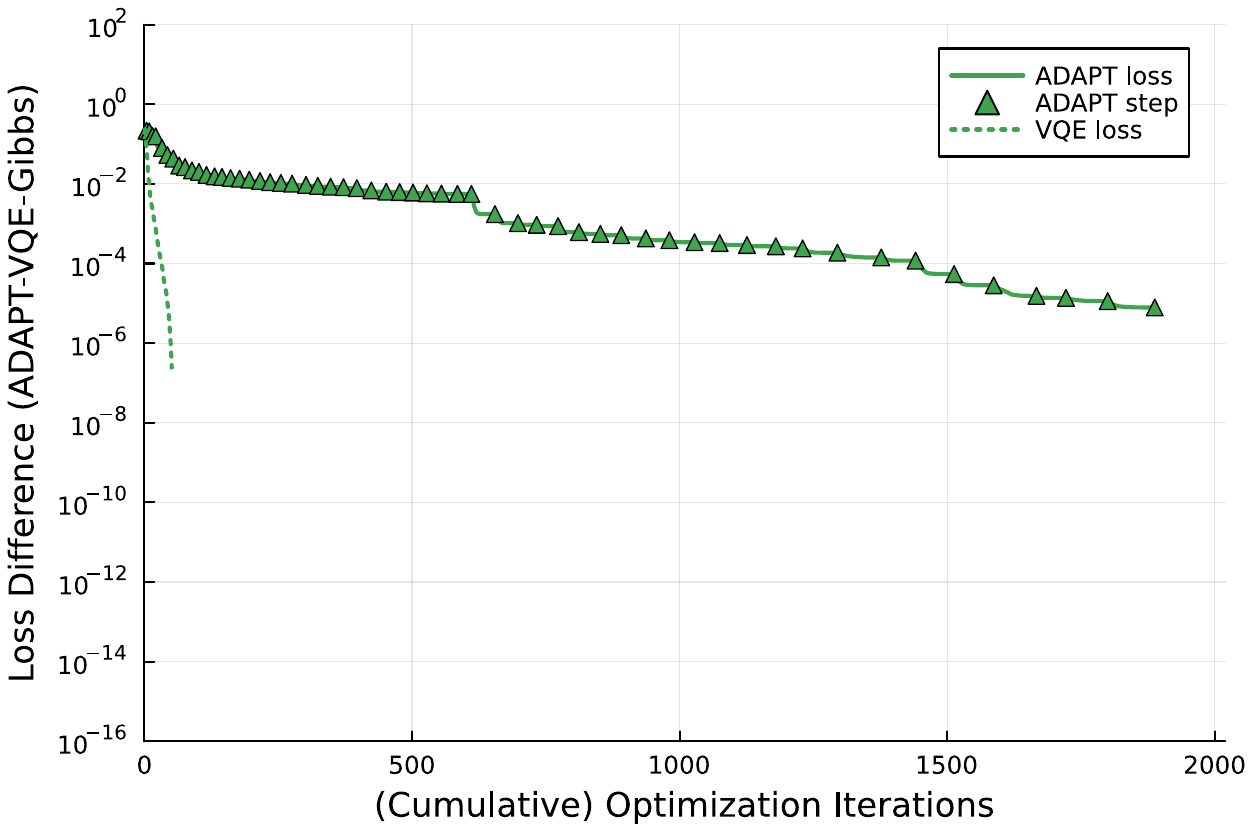}
    \caption{Loss curves to train a random two-local thermal state on 3 system qubits and 3 ancillae, using the maximal Renyi divergence (\ref{renyi-loss-fn-reversed}), the overlap (\ref{overlap-adapt-loss-fn}), and the modified objective in ADAPT-VQE-Gibbs (\ref{gibbs-loss-fn}) as cost functions. Note that the latter has its minimum at a non-zero value, so we plot the difference between the loss of the trial state and the target state. Markers indicate where individual optimizations completed and a new Pauli rotation was added in the ADAPT runs. The dotted lines give the loss curve when running a single VQE optimization on an ansatz consisting of all 153 Pauli rotations.}
    
    \label{fig:adapt_v_vqe}
\end{figure}

Like the other adaptive methods, Rényi-ADAPT successfully learned the target state (Fig. \ref{fig:adapt_v_vqe}). Because the VQE ans\"atze start out sufficiently expressive to reach the target state with just one optimization, they required drastically fewer iterations to converge compared to the adaptive methods, implying the fixed-structure approach would require drastically fewer circuit evaluations in a quantum experiment. In contrast, adaptive methods achieved excellent loss values with at least a 60\% reduction in parameter count (Table \ref{tab:adapt_v_vqe}) compared to the VQE ansätze. This translates to a commensurate reduction of circuit depth in a quantum experiment, making the adaptive approach the preferred method for devices with limited coherence times.

\begin{table}[]
    \centering
    \caption{Parameter counts of each algorithm}
    \label{tab:adapt_v_vqe}
    \begin{tabular}{lcc}
    \hline
        \textbf{Algorithm} & \textbf{Parameters} \\\hline
        VQE & 153 \\
        Overlap-ADAPT & \textbf{52} \\
        ADAPT-VQE-Gibbs & 57 \\
        Rényi-ADAPT & 61 \\
    \end{tabular}
\end{table}

\subsection{Increasing system size}

Next, we investigate the robustness of our method for larger system sizes. Similarly to the previous experiment, each algorithm was tested in 20 random thermal states, where the number of visible qubits now varied in the range $n \in [1, 4]$. For each method, we recorded the worst-case infidelity over all instances after each ADAPT iteration, enabling a direct comparison of compactness between the three loss functions.

\begin{figure}
    \includegraphics[width=\columnwidth]{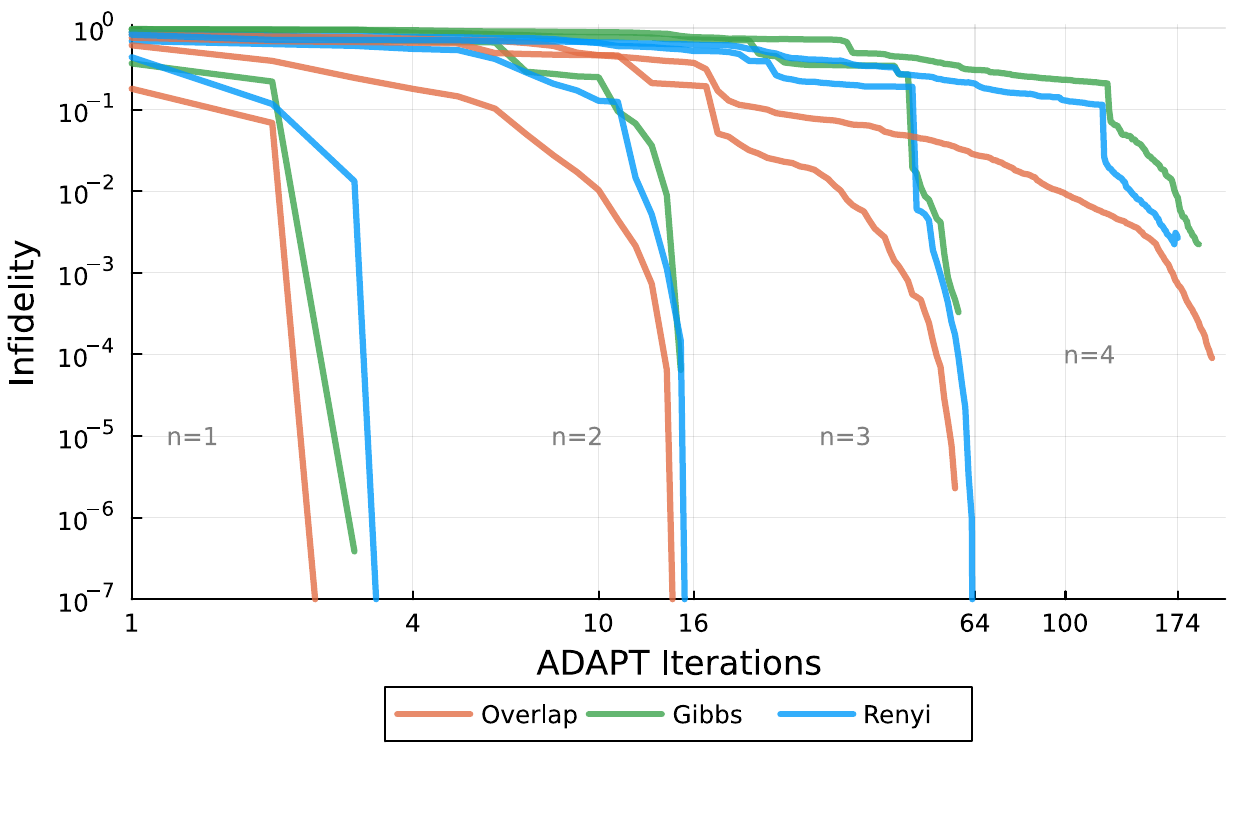}
    \caption{ADAPT convergence to train a random two-local thermal state, for each choice of loss function. Different line styles show results for different system sizes $n$. The results shown are worst-case infidelities collected from up to 20 trials of randomly sampled target states and reference states (due to time constraints, we completed only one successful trial with $n=4$ from each method).}
    \label{fig:adapt_methods}
\end{figure}

For all system sizes, overlap-ADAPT achieved a given infidelity threshold with the fewest parameters, but performance was competitive among the loss functions (Fig. \ref{fig:adapt_methods}). At the largest system size of $n = 4$, ADAPT-VQE-Gibbs and Rényi-ADAPT stopped near infidelities of $10^{-3}$ while overlap-ADAPT reached $10^{-4}$. However, we were only able to run 1 trial at this size due to computational constraints, and so this behavior may not be representative. Interestingly, all methods required exponentially deeper circuits as $n$ increased due to the larger Hilbert space dimension and the locality of the operator pool.
These simulations suggest that the Rényi divergence does not provide an advantage in terms of circuit depth.

It is worth mentioning here that the loss functions for R\'enyi divergence $\widetilde{D}_2(\rho\|\sigma(\theta))$ (\ref{renyi-loss-fn-reversed}) and overlap $\mathcal{L}_O(\rho, \sigma(\theta))$ (\ref{overlap-adapt-loss-fn}) were evaluated using the exact thermal state $\rho$, while the loss of ADAPT-VQE-Gibbs $C(\sigma(\theta))$ (\ref{gibbs-loss-fn}) was calculated using a truncated Taylor expansion of the thermal state, thus limiting the fidelity it could achieve (although it still achieves $> 99.8\%$ fidelity for the $n=4$ trial shown). This approximation was intentional; in most quantum machine learning applications, the target state $\rho$ is known, whereas in ADAPT-VQE-Gibbs, which was designed for quantum chemical and condensed matter systems, $\rho$ is the desired output for which \textit{a priori} estimates are not guaranteed, necessitating the need for an approximation.

\subsection{Gradient results}

The unbounded nature of the maximal Rényi divergence lends it resilience to barren plateaus, as discussed by Kieferova et al. \cite{kieferova2021quantum}, providing better trainability for variational circuits. We investigated this claim by analyzing the gradients of the loss functions.

In a similar setup as before, using 20 random combinations of the target state $\rho$ and the reference state $\sigma_0$, we calculated the initial pool gradients used to select the first operator. We denote the magnitude of the largest gradient as $||g||_\infty$, which we refer to as the $L_\infty$ norm of the pool gradient. Because this did not require the full ADAPT procedure, we were able to run experiments on a system size of $n=6$ (12 total qubits, with $n_H = n_V)$. We expected that barren plateaus would manifest as an exponential decay of $||g||_\infty$ as $n$ increased, and therefore hypothesized that the Rényi divergence would not decay.


\begin{figure}
    \includegraphics[width=\columnwidth]{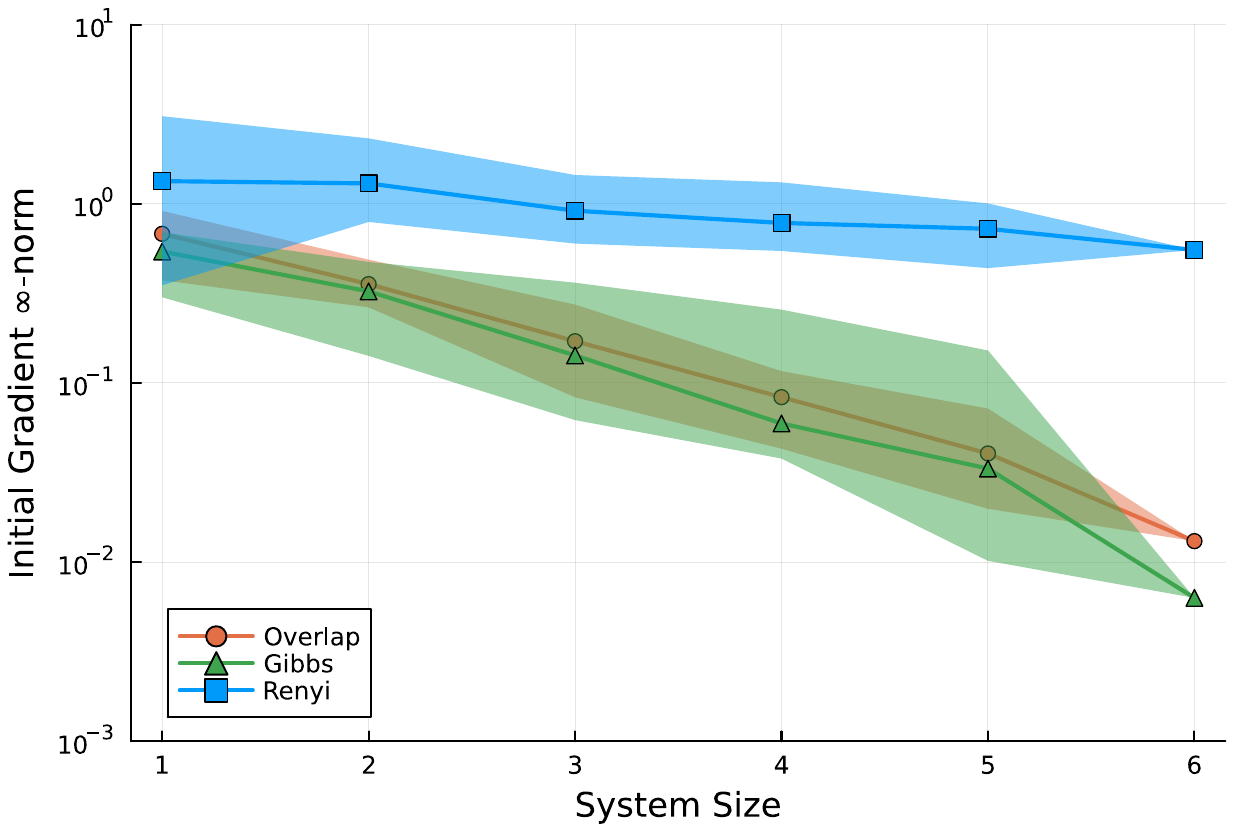}
    \caption{Largest pool gradient in the first ADAPT selection, as a function of system size $n$. Error bars show the full range of values from up to 20 trials of randomly sampled target states and reference states. As $n$ increases, all pool gradients for the overlap and for ADAPT-VQE-Gibbs vanish exponentially, indicating the presence of barren plateaus. For large enough systems, the gradients will be too small to resolve the optimal choice of operator, and ADAPT will fail to start. However, gradients for the Rényi divergence are nearly constant.}
    \label{fig:first_steps}
\end{figure}

The results are shown in Fig. \ref{fig:first_steps} with the lines indicating the median  of the initial pool gradient $\infty$-norm over the trials and the shaded regions depicting the full range of values. The Rényi divergence exhibited larger, nearly constant gradients as the system size $n$ increased, while the overlap and Gibbs loss functions decayed exponentially at rates similar to each other. 


Table \ref{tab:fitted_exp} contains exponential functions fitted to the medians of each loss function with respect to $n$. While Rényi-ADAPT still suffers exponential decay in the large $n$ limit, the decay is significantly slower. To illustrate by how much slower, let us imagine that the gradient measurement has a resolution of $10^{-5}$, such that when all operators have gradients below this threshold (i.e. $||g||_\infty < 10^{-5}$), the ADAPT protocol cannot meaningfully begin. We denote this in Table \ref{tab:fitted_exp} as the ``predicted failure". Using the loss functions for either overlap-ADAPT or ADAPT-VQE-Gibbs, the algorithm would fail to start at $n \approx 15$ qubits (16 and 14, respectively), while with the Rényi divergence, this threshold is not reached until $n = 67$, a 4x improvement.
For our lower-precision criteria of $||g||_\infty = 10^{-3}$, we expect that overlap-ADAPT and ADAPT-VQE-Gibbs will become untrainable at system sizes of $n \approx 10$. Note that our simulations ($n \leq 6)$ are well below this regime, such that this advantage of Rényi-ADAPT is not reflected in the experiments whose results are shown in Figs. \ref{fig:adapt_v_vqe} and \ref{fig:adapt_methods}.


\begin{table}[]
    \centering
    \caption{Decay of initial pool gradient}
    \label{tab:fitted_exp}
    \begin{tabular}{ccc}
        \hline
        \textbf{Loss} & \textbf{Fitted curve} & \textbf{Predicted failure at $10^{-5}$} \\\hline
        Overlap & $||g||_\infty = 1.644 \times 2.162^{-n}$ & $n=16$\\
        Gibbs & $||g||_\infty = 1.648 \times 2.354^{-n}$ & $n=14$\\
        Rényi & $||g||_\infty = 1.676 \times 1.198^{-n}$ & $n=67$
    \end{tabular}
\end{table}

\subsection{Initial state}


The choice of reference state is important for the performance of ADAPT, with good reference states being close to the target~\cite{Feniou_2023}. For poor reference states, ADAPT may require many more operators, removing its compact-circuit advantage, or even fail to converge altogether. Some examples of good reference states include the Hartree-Fock state for the ground state of molecular electronic structure problems \cite{Grimsley2019a} and the Néel antiferromagnetic state for the ground state of lattice spin systems~\cite{OperatorTiling2024}. For overlap-ADAPT and Rényi-ADAPT, the ideal reference state is, of course, $\sigma_0 = \rho$.

An interesting possibility of the Rényi divergence is that it could avoid concentrating (or flattening out) far away from $\rho$. If true, this would improve the trainability of circuits and reduce the need to choose a good reference state, which would be helpful when one is either hard to produce or completely unknown for the given problem.

\begin{figure}
    \includegraphics[width=\columnwidth]{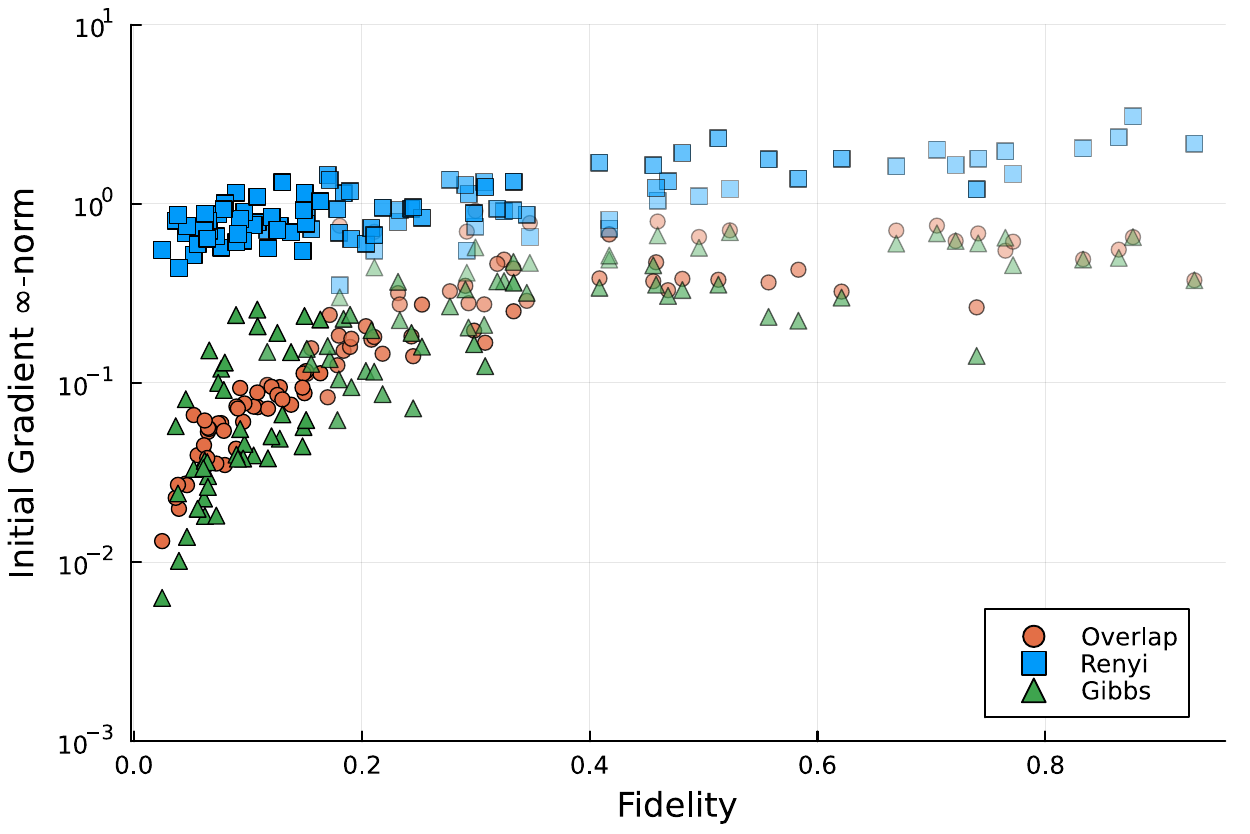}
    \caption{Largest pool gradient in the first ADAPT selection, as a function of fidelity between the reference state $\sigma_0$ and the target thermal state $\rho$. Markers for smaller systems are plotted with more transparency. When the fidelity starts very small, its gradient is also very small, indicating the presence of a barren plateau in the cost-function landscape far from the target state. By contrast, the Renyi divergence exhibits large gradients throughout.}
    \label{fig:distance}
\end{figure}

After processing the data in Fig. \ref{fig:adapt_methods}, we calculated the fidelities $F(\rho, \sigma_0)$ between the random target and reference states. The initial pool gradient $||g||_\infty$ is shown in Fig. \ref{fig:distance} as a function of $F$. We observed that further away from $\rho$, the gradient landscapes for ADAPT-VQE-Gibbs and overlap-ADAPT decayed exponentially, suggesting that trainability could become an issue. In contrast, Rényi-ADAPT again demonstrated favorable near-constant gradients (indicative of the loss landscape not flattening), supporting the idea that it functions effectively with a wider range of reference states.


\subsection{Completion}

\begin{figure}
    \centering
    \includegraphics[width=\columnwidth]{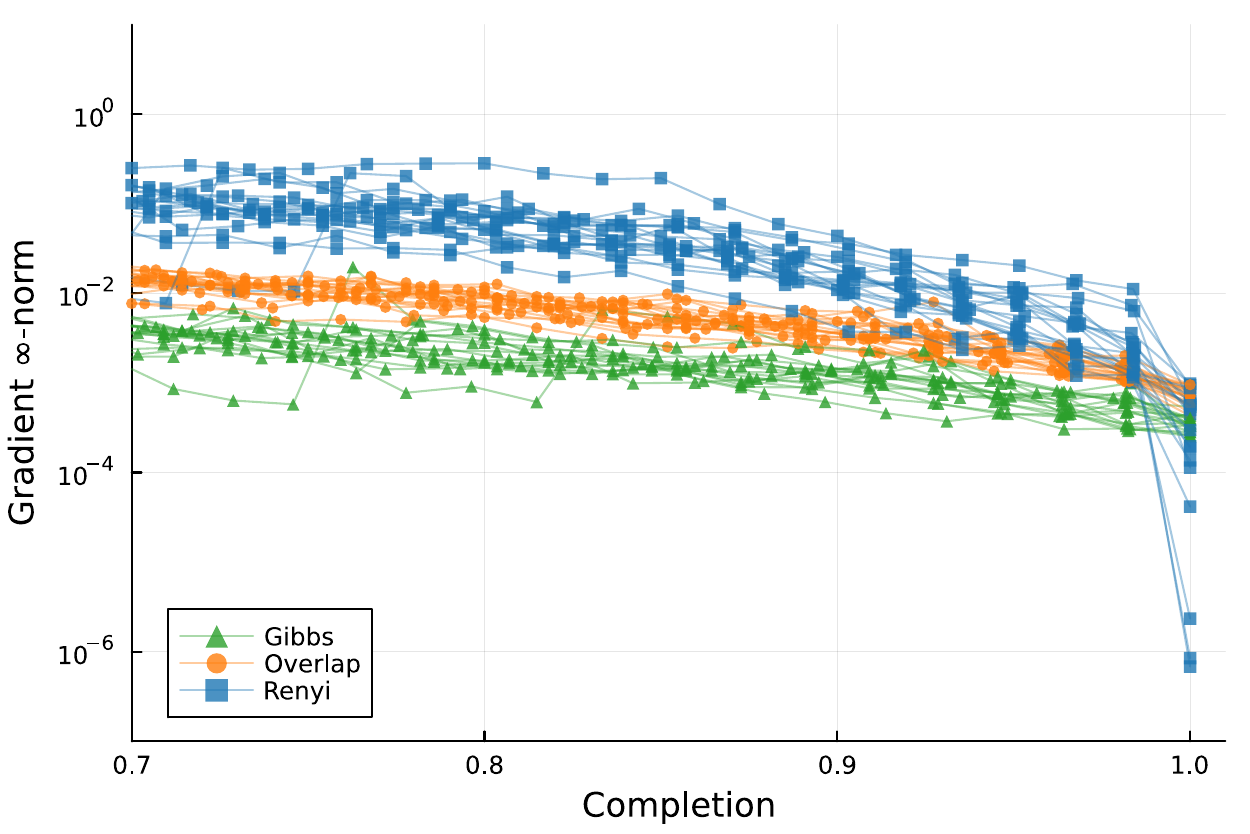}
    \caption{Largest pool gradient element over the course of twenty ADAPT runs for $n = 3$. Completion is calculated per run as the number of ansatz parameters divided by the final ansatz length, which was defined to terminate once the pool gradient fell below $10^{-3}$. On average, gradients decline exponentially (linear on the log plot) as ADAPT progresses. However, once an ansatz is sufficiently expressive, no further improvements can be made and the gradients drop off rapidly. For a threshold of $\varepsilon=10^{-3}$, only Rényi-ADAPT reaches this regime (albeit with a longer ansatz, as can be seen Figs.~\ref{fig:adapt_v_vqe} and~\ref{fig:adapt_methods}).}
    \label{fig:completion}
\end{figure}

ADAPT-VQE experiments require a specific choice of the value $\varepsilon$ to take as the gradient norm threshold.
In this work, we adopted the somewhat arbitrary choice of $\varepsilon=1e-3$.
To assess the degree to which this choice was appropriate, we examine the $L_\infty$ norm of the pool gradient $||g||_\infty$ for each ADAPT iteration, for all 20 trials where $n=3$ qubits.
As a general trend, pool gradients tend to fall exponentially as each parameter increases the expressivity of the ansatz.
However, when no further improvements can be made to the trial state, all pool gradients will rapidly drop to zero.
To highlight this behavior, we plot in Fig.~\ref{fig:completion} the $L_\infty$ norm of the pool gradient as a function of the percent completion of the ADAPT run, where a ``completed'' run is one where the final $||g||_\infty<\varepsilon$.

Our results show that, on average, $||g||_\infty$ values for both ADAPT-VQE-Gibbs and overlap-ADAPT decay only exponentially all the way through the $||g||_\infty<\varepsilon$ stopping condition.
Meanwhile, the Rényi-ADAPT curves exhibit super-exponential decay roughly 80\% of the way to the same stopping condition, followed by several runs dropping from $||g||_\infty>1e-3$ to $||g||_\infty<1e-4$ in a single ADAPT iteration.
We reiterate that Rényi-ADAPT tended to require deeper circuits to attain this convergence (as observed in Figs.~\ref{fig:adapt_v_vqe} and~\ref{fig:adapt_methods}), so this observation does not necessarily indicate an advantage for Rényi-ADAPT, especially in terms of near-term viability.
However, this does suggest that Rényi-ADAPT requires a lower minimum resolution when measuring pool gradients to attain a region of strong convergence, implying that each pool operator could be selected in Rényi-ADAPT using a lower shot count than in the other methods, helping to offset the major cost of measurement overhead incurred by the ADAPT protocol.


\section{Discussion}

In this work, we characterize Rényi-ADAPT as a method resistant to barren plateaus for preparing thermal states. It combines the trainability advantages of the maximal Rényi divergence \cite{kieferova2021quantum} with the adaptive optimization procedure of ADAPT-VQE \cite{Grimsley2019a}. We did not observe the gradients obtained in Rényi-ADAPT to decay in our experiments. We conclude with a discussion of our results and how they relate to the existing literature on barren plateaus.

A recent review of barren plateaus by Larocca et al. \cite{larocca2024review} summarizes the types and origins of barren plateaus. They can arise from highly expressive circuits, poor choice of initial state, global measurements, and noise. The relationship between these is studied for 2-designs in \cite{ragone2023unified, fontana2023adjoint}.

{\bf Expressiveness} By iteratively constructing circuits, ADAPT balances the trade-off between constructing a sufficiently expressive circuit to reach the objective minimum and overparameterizing the circuit, which would induce barren plateaus.

Our results in Figs. \ref{fig:adapt_v_vqe} and \ref{fig:adapt_methods} show overlap-ADAPT producing the most compact circuits in terms of parameter count. On devices where producing the most compact circuits is a priority (like near-term devices with limited coherence times), we surmise that the overlap would be the best choice of the algorithms we tested, provided a suitable reference state.

Note that the expressiveness of the circuit is characterized by the size of its dynamical Lie algebra (DLA) \cite{larocca2024review, ragone2023unified, fontana2023adjoint}. Since ADAPT can, in principle, append a previously-added operator (which would not increase the DLA), the DLA may be smaller than expected by assuming unique operators for each parameter. However, similar to the existing ADAPT literature \cite{Grimsley2019a, OperatorTiling2024, warren2022adaptive, Feniou_2023}, we use parameter counts as a proxy for circuit expressiveness and leave the quantification of DLA to future work.


{\bf Measurement} The compactness of ADAPT comes at the considerable expense of measuring the operator pool at each iteration. The 1- and 2-local pool (which has $3(n_V + n_H) + 9\cdot \binom{n_V+n_H}{2}$ operators) we used is complete
. However, by using these small, local operators, constructing a large unitary requires many more iterations. We could alternatively choose a {\em minimal} operator pool, which has been shown to have a number of operators linear in the number of qubits. \cite{Tang2021,Shkolnikov_2023} This minimizes the measurement overhead in each ADAPT iteration, at the cost of requiring an even longer circuit to reach arbitrary states.

On the other hand, by expanding the pool to include all $k$-local Paulis, we can reduce the number of parameters required in the circuit, at the cost of significantly more measurements required to measure the full pool gradient. According to current barren plateau research, the increased globality of the measurements could be problematic, suggesting there is a trade-off between including larger pool operators and trainability. We did not investigate this in our work.



{\bf Initial State} Choosing a random reference state increases the chances of a barren plateau, following the curse of dimensionality \cite{larocca2024review}. As detailed in Appendix~\ref{computational-details}, we had to choose a random reference state to allow ADAPT to construct mixed states. 

Our experiments show that Rényi-ADAPT has larger gradients for our reference states both with respect to increasing system size (Fig. \ref{fig:first_steps}) and larger distances from the objective (Fig. \ref{fig:distance}). Therefore, at least in the absence of a good reference state, overlap-ADAPT and ADAPT-VQE-Gibbs should have to make exponentially more measurements of the gradient, or they might possibly not start at all, a reason to favor Rényi-ADAPT.


As we used a state vector simulator with infinite shots, this was not shown in our experiments. Additionally, our simulations did not reach the scale where we anticipate overlap-ADAPT and ADAPT-VQE-Gibbs to start below the convergence criteria.

{\bf Noise} Noise-induced barren plateaus are perhaps the most significant obstacle to leveraging near-term quantum devices that lack error correction. Longer circuits are generally noisier, which is the original motivation behind ADAPT-VQE \cite{Grimsley2019a}.

We did not test with a noisy quantum simulator, which we leave to future work. Our noiseless results show overlap-ADAPT producing more compact circuits, which one would expect to have less noise. However, as noise concentrates the loss function, this restricts the space of good reference states, which we found to be a potential obstacle to the scalability of overlap-ADAPT. Investigating this interplay would be interesting. Additionally, since Rényi-ADAPT was more trainable (initially) with poor reference states, one could imagine that a hybrid algorithm where Rényi-ADAPT prepares a reference state for overlap-ADAPT might perform well, combining the advantages of each algorithm we observed.

\section{Data Availability}

The code and data used for the simulations and figures can be found at \url{https://github.com/kmsherbertvt/RenyiADAPT}.

\section*{Acknowledgment}
The authors would like to thank Ada Warren for her helpful comments and discussion on the ADAPT-VQE-Gibbs code.
This research was supported by the DOE Office of Science, National Quantum Information Science Research Centers, Co-design Center for Quantum Advantage (C2QA) under contract number DE-SC0012704.

\bibliographystyle{unsrt}  
\bibliography{refs}

\begin{thebibliography}{10}

\bibitem{Peruzzo2014}
Alberto Peruzzo, Jarrod McClean, Peter Shadbolt, Man-Hong Yung, Xiao-Qi Zhou, Peter~J. Love, Al{\'a}n Aspuru-Guzik, and Jeremy~L. O'Brien.
\newblock A variational eigenvalue solver on a photonic quantum processor.
\newblock {\em Nature Communications}, 5(1):4213, Jul 2014.

\bibitem{farhi2014quantum}
Edward Farhi, Jeffrey Goldstone, and Sam Gutmann.
\newblock A quantum approximate optimization algorithm.
\newblock {\em arXiv preprint arXiv:1411.4028}, 2014.

\bibitem{havlivcek2019supervised}
Vojt{\v{e}}ch Havl{\'\i}{\v{c}}ek, Antonio~D C{\'o}rcoles, Kristan Temme, Aram~W Harrow, Abhinav Kandala, Jerry~M Chow, and Jay~M Gambetta.
\newblock Supervised learning with quantum-enhanced feature spaces.
\newblock {\em Nature}, 567(7747):209--212, 2019.

\bibitem{schuld2020circuit}
Maria Schuld, Alex Bocharov, Krysta~M Svore, and Nathan Wiebe.
\newblock Circuit-centric quantum classifiers.
\newblock {\em Physical Review A}, 101(3):032308, 2020.

\bibitem{ceroni2022generating}
Jack Ceroni, Torin~F Stetina, Maria Kieferova, Carlos~Ortiz Marrero, Juan~Miguel Arrazola, and Nathan Wiebe.
\newblock Generating approximate ground states of molecules using quantum machine learning.
\newblock {\em arXiv preprint arXiv:2210.05489}, 2022.

\bibitem{Feniou_2023}
César Feniou, Muhammad Hassan, Diata Traoré, Emmanuel Giner, Yvon Maday, and Jean-Philip Piquemal.
\newblock Overlap-adapt-vqe: practical quantum chemistry on quantum computers via overlap-guided compact ansätze.
\newblock {\em Communications Physics}, 6(1), July 2023.

\bibitem{larocca2024review}
Martin Larocca, Supanut Thanasilp, Samson Wang, Kunal Sharma, Jacob Biamonte, Patrick~J. Coles, Lukasz Cincio, Jarrod~R. McClean, Zoë Holmes, and M.~Cerezo.
\newblock A review of barren plateaus in variational quantum computing, 2024.

\bibitem{grant2019initialization}
Edward Grant, Leonard Wossnig, Mateusz Ostaszewski, and Marcello Benedetti.
\newblock An initialization strategy for addressing barren plateaus in parametrized quantum circuits.
\newblock {\em Quantum}, 3:214, 2019.

\bibitem{skolik2021layerwise}
Andrea Skolik, Jarrod~R McClean, Masoud Mohseni, Patrick van~der Smagt, and Martin Leib.
\newblock Layerwise learning for quantum neural networks.
\newblock {\em Quantum Machine Intelligence}, 3(1):1--11, 2021.

\bibitem{pesah2020absence}
Arthur Pesah, M~Cerezo, Samson Wang, Tyler Volkoff, Andrew~T Sornborger, and Patrick~J Coles.
\newblock Absence of barren plateaus in quantum convolutional neural networks.
\newblock {\em arXiv preprint arXiv:2011.02966}, 2020.

\bibitem{cerezo2020cost}
M~Cerezo, Akira Sone, Tyler Volkoff, Lukasz Cincio, and Patrick~J Coles.
\newblock Cost-function-dependent barren plateaus in shallow quantum neural networks.
\newblock {\em arXiv preprint arXiv:2001.00550}, 2020.

\bibitem{sharma2020trainability}
Kunal Sharma, M~Cerezo, Lukasz Cincio, and Patrick~J Coles.
\newblock Trainability of dissipative perceptron-based quantum neural networks.
\newblock {\em arXiv preprint arXiv:2005.12458}, 2020.

\bibitem{kieferova2021quantum}
Maria Kieferova, Carlos Ortiz~Marrero, and Nathan Wiebe.
\newblock Quantum generative training using r\'enyi divergences.
\newblock {\em arXiv preprint arXiv:2106.09567}, 2021.

\bibitem{warren2022adaptive}
Ada Warren, Linghua Zhu, Nicholas~J. Mayhall, Edwin Barnes, and Sophia~E. Economou.
\newblock Adaptive variational algorithms for quantum gibbs state preparation, 2022.

\bibitem{Cerezo2021}
M.~Cerezo, Andrew Arrasmith, Ryan Babbush, Simon~C. Benjamin, Suguru Endo, Keisuke Fujii, Jarrod~R. McClean, Kosuke Mitarai, Xiao Yuan, Lukasz Cincio, and Patrick~J. Coles.
\newblock Variational quantum algorithms.
\newblock {\em Nature Reviews Physics}, 3(9):625--644, Sep 2021.

\bibitem{Tilly2022}
Jules Tilly, Hongxiang Chen, Shuxiang Cao, Dario Picozzi, Kanav Setia, Ying Li, Edward Grant, Leonard Wossnig, Ivan Rungger, George~H. Booth, and Jonathan Tennyson.
\newblock The variational quantum eigensolver: A review of methods and best practices.
\newblock {\em Physics Reports}, 986:1--128, 2022.
\newblock The Variational Quantum Eigensolver: a review of methods and best practices.

\bibitem{Bharti2022}
Kishor Bharti, Alba Cervera-Lierta, Thi~Ha Kyaw, Tobias Haug, Sumner Alperin-Lea, Abhinav Anand, Matthias Degroote, Hermanni Heimonen, Jakob~S. Kottmann, Tim Menke, Wai-Keong Mok, Sukin Sim, Leong-Chuan Kwek, and Al\'an Aspuru-Guzik.
\newblock Noisy intermediate-scale quantum algorithms.
\newblock {\em Rev. Mod. Phys.}, 94:015004, Feb 2022.

\bibitem{Grimsley2019a}
Harper~R. Grimsley, Sophia~E. Economou, Edwin Barnes, and Nicholas~J. Mayhall.
\newblock An adaptive variational algorithm for exact molecular simulations on a quantum computer.
\newblock {\em Nature Communications}, 10(1):1--9, July 2019.

\bibitem{Zhang2021}
Feng Zhang, Niladri Gomes, Yongxin Yao, Peter~P. Orth, and Thomas Iadecola.
\newblock Adaptive variational quantum eigensolvers for highly excited states.
\newblock {\em Phys. Rev. B}, 104:075159, Aug 2021.

\bibitem{Gomes2021}
Niladri Gomes, Anirban Mukherjee, Feng Zhang, Thomas Iadecola, Cai-Zhuang Wang, Kai-Ming Ho, Peter~P. Orth, and Yong-Xin Yao.
\newblock Adaptive variational quantum imaginary time evolution approach for ground state preparation.
\newblock {\em Advanced Quantum Technologies}, 4(12):2100114, 2021.

\bibitem{Yao2021}
Yong-Xin Yao, Niladri Gomes, Feng Zhang, Cai-Zhuang Wang, Kai-Ming Ho, Thomas Iadecola, and Peter~P. Orth.
\newblock Adaptive variational quantum dynamics simulations.
\newblock {\em PRX Quantum}, 2:030307, Jul 2021.

\bibitem{LZhu2020}
Linghua Zhu, Ho~Lun Tang, George~S. Barron, F.~A. Calderon-Vargas, Nicholas~J. Mayhall, Edwin Barnes, and Sophia~E. Economou.
\newblock Adaptive quantum approximate optimization algorithm for solving combinatorial problems on a quantum computer.
\newblock {\em Phys. Rev. Res.}, 4:033029, Jul 2022.

\bibitem{chen2023entanglement}
Yanzhu Chen, Linghua Zhu, Chenxu Liu, Nicholas~J. Mayhall, Edwin Barnes, and Sophia~E. Economou.
\newblock How much entanglement do quantum optimization algorithms require?, 2023.

\bibitem{wilde2014strong}
Mark~M Wilde, Andreas Winter, and Dong Yang.
\newblock Strong converse for the classical capacity of entanglement-breaking and hadamard channels via a sandwiched r{\'e}nyi relative entropy.
\newblock {\em Communications in Mathematical Physics}, 331(2):593--622, 2014.

\bibitem{muller2013quantum}
Martin M{\"u}ller-Lennert, Fr{\'e}d{\'e}ric Dupuis, Oleg Szehr, Serge Fehr, and Marco Tomamichel.
\newblock On quantum r{\'e}nyi entropies: A new generalization and some properties.
\newblock {\em Journal of Mathematical Physics}, 54(12):122203, 2013.

\bibitem{petz1986quasi}
D{\'e}nes Petz.
\newblock Quasi-entropies for finite quantum systems.
\newblock {\em Reports on mathematical physics}, 23(1):57--65, 1986.

\bibitem{renyi}
Tim Van~Erven and Peter Harremos.
\newblock R{\'e}nyi divergence and kullback-leibler divergence.
\newblock {\em IEEE Transactions on Information Theory}, 60(7):3797--3820, 2014.

\bibitem{optim}
Patrick~Kofod Mogensen and Asbj{\o}rn~Nilsen Riseth.
\newblock Optim: A mathematical optimization package for {Julia}.
\newblock {\em Journal of Open Source Software}, 3(24):615, 2018.

\bibitem{julia}
Jeff Bezanson, Alan Edelman, Stefan Karpinski, and Viral~B. Shah.
\newblock Julia: {A} {Fresh} {Approach} to {Numerical} {Computing}.
\newblock {\em SIAM Review}, 59(1):65--98, January 2017.
\newblock Publisher: Society for Industrial and Applied Mathematics.

\bibitem{OperatorTiling2024}
John~S. Van~Dyke, Karunya Shirali, George~S. Barron, Nicholas~J. Mayhall, Edwin Barnes, and Sophia~E. Economou.
\newblock Scaling adaptive quantum simulation algorithms via operator pool tiling.
\newblock {\em Phys. Rev. Res.}, 6:L012030, Feb 2024.

\bibitem{ragone2023unified}
Michael Ragone, B.~Bakalov, Frédéric Sauvage, Alexander~F. Kemper, Carlos~Ortiz Marrero, Mart{\' i}n Larocca, and M.~Cerezo.
\newblock A {Unified} {Theory} of {Barren} {Plateaus} for {Deep} {Parametrized} {Quantum} {Circuits}.
\newblock {\em arXiv}, 2023.

\bibitem{fontana2023adjoint}
Enrico Fontana, Dylan Herman, Shouvanik Chakrabarti, Niraj Kumar, Romina Yalovetzky, Jamie Heredge, Shree~Hari Sureshbabu, and Marco Pistoia.
\newblock The {Adjoint} {Is} {All} {You} {Need}: Characterizing {Barren} {Plateaus} in {Quantum} {Ansätze}.
\newblock {\em arXiv}, 2023.

\bibitem{Tang2021}
Ho~Lun Tang, V.O. Shkolnikov, George~S. Barron, Harper~R. Grimsley, Nicholas~J. Mayhall, Edwin Barnes, and Sophia~E. Economou.
\newblock Qubit-adapt-vqe: An adaptive algorithm for constructing hardware-efficient ans\"atze on a quantum processor.
\newblock {\em PRX Quantum}, 2:020310, Apr 2021.

\bibitem{Shkolnikov_2023}
V.~O. Shkolnikov, Nicholas~J. Mayhall, Sophia~E. Economou, and Edwin Barnes.
\newblock Avoiding symmetry roadblocks and minimizing the measurement overhead of adaptive variational quantum eigensolvers.
\newblock {\em Quantum}, 7:1040, June 2023.

\end{thebibliography}

\newpage
\appendix

\subsection{Hyperparameters}\label{computational-details}

We give the hyperparameters of our experiments in Table \ref{tab:hyperparameters}. $\mathcal{P}_k^{(n)}$ denotes the set of all $k$-local Pauli operators that act on $n$ qubits. The gradient convergence expressions refer to the gradient of the operator pool.

\begin{table}[]
    \centering
    \begin{tabular}{lc}
    \hline
        \textbf{Hyperparameter} & \textbf{Value}\\\hline
        Target qubits $n$ & 1-6 \\
        Hidden qubits $n_H$ & Set to $n$\\\hline

        Overlap convergence $||\nabla F||_\infty$ & $10^{-3}$ \\
        Gibbs convergence $||\nabla C||_\infty$ & $10^{-3}$ \\
        Rényi convergence $||\nabla \tilde{D}_2||_\infty$ & $10^{-3}$ \\\hline

        Operator pool & $\mathcal{P}_1^{(2n)} \cup \mathcal{P}_2^{(2n)}$
    \end{tabular}
    \caption{Experiment hyperparameters}
    \label{tab:hyperparameters}
\end{table}

The Hamiltonians used in the experiments were generated by assigning random coefficients drawn from a normal distribution with zero mean and unit variance to one- and two-local Pauli operators on the $n_V$ visible qubits. To ensure that the energy scale remained the same with different system sizes, the Hamiltonians $\Hat{H} = \sum_i c_i P_i$ (where each $P_i$ is one- or two-local) were subsequently normalized such that $c_i \rightarrow \frac{c_i}{||\mathbf{c}||}$. Twenty such Hamiltonians were generated to sample various two-local models. The operator pool consisted of all one- and two-local Pauli operators on the total number of qubits $n_V + n_H$.

The reference states used were produced by generating states that are partially-entangled between the visible and hidden qubits. As discussed in previous work~\cite{warren2022adaptive}, reference states without any entanglement between the visible and hidden qubits maximize the purity $\text{Tr}(\rho^2)$ of $\rho = \text{Tr}_A |\psi\rangle\langle\psi|$. Non-zero gradients of the ADAPT-VQE-Gibbs objective function are then obtained only from the first term in (\ref{gibbs-loss-fn}); the structure of the pool operators then implies that the operator with the largest gradient will always be local to the visible system qubits, without the ability to generate entanglement between it and the hidden system. This would result in the true Gibbs state (a mixed state) being unattainable. A similar argument can be made against the use of maximally mixed reference states. Correspondingly, for Rényi-ADAPT, it was found that for product reference states, individual Pauli operators were incapable of generating entanglement between the visible and hidden qubits, and that operators consisting of linear combinations of diagonal and off-diagonal Pauli operators (`diagonal' here refers to operators whose eigenvectors the product reference is composed of) would be necessary in order for such entanglement to be created.

In order to overcome the constraint on the pool operators arising from using product states, partially-entangled reference states were generated by first applying random single-qubit $R_y$ rotations to each qubit in the state $|0\rangle^{\otimes n_V+n_H}$, with the angle being drawn from $\left[ - \pi, \pi\right) $, followed by a layer of CNOT gates between each visible-hidden qubit pair, with the control being on the hidden qubits. Twenty such reference states were generated to sample various initializations. We leave the investigation of operator pools consisting of linear combinations of Pauli operators to future work.

The convergence criterion for ADAPT was set such that the calculation was ended when the gradient norm of the pool operators was below $1e-3$.

\subsection{Gradients}\label{gradients}
In order to select operators and conduct gradient descent, it is essential to have a formula for gradients that can be measured in a quantum computer. Below we provide a discussion of the closed form expression of the gradients of each loss function that we considered in our work:


\subsubsection{Overlap-ADAPT}

The expression for the pool operator gradients at step $k+1$ may be written as~\cite{Feniou_2023}
\begin{multline}\label{overlap-gradient}
     \frac{\partial}{\partial \theta_{k+1}} \left. \langle \Psi_{\rm target} | e^{-i\theta_{k+1} A_{k+1}} \psi_{k}(\vect{\theta}^{(k)})\rangle \right|_{\theta_{k+1}=0} \\ = \langle\Psi_{\rm target} | i A_{k+1} \psi_{k}(\vect{\theta}^{(k)})\rangle.
\end{multline}

\subsubsection{ADAPT-VQE-Gibbs}
In the case where the ansatz is composed of operators $e^{-i\theta G}$ whose Hermitian generators $G$ have exactly two distinct eigenvalues $\epsilon_0$ and $\epsilon_1$, the exact gradients of the objective function can be efficiently obtained using the parameter-shift rule. Defining an auxiliary function 
\begin{equation}
    \tilde{C}(\theta,\phi) = - \text{Tr}(\rho_G \sigma(\theta)) + \frac{1}{2} \text{Tr}(\sigma(\theta) \sigma(\phi)),
\end{equation}
the gradient of $C(\sigma(\theta))$ may be written as
\begin{equation}
    \frac{\partial}{\partial \theta} C(\sigma(\theta)) = r \left[ \tilde{C}(\theta + \frac{\pi}{4r},\theta) - \tilde{C}(\theta - \frac{\pi}{4r},\theta) \right],
\end{equation}
where $r = \frac{1}{2}(\epsilon_1 - \epsilon_0)$. 

\subsubsection{R\'enyi-ADAPT}
This divergence has closed-form gradients that can be efficiently computed on a Quantum Computer. Define a parameterized quantum circuit by $\sigma_v := \text{Tr}_h\left[ \Pi_{j=1}^N e^{-iH_j\theta_j} \ket{\psi_{\rm ref}}\bra{\psi_{\rm ref}} \Pi_{j=N}^1 e^{iH_j\theta_j} \right]$, where we measure the output density by taking a partial trace, denoted by $\text{Tr}_h$, over a hidden subspace of some set of ancilla qubits. In this case, the gradients of equation \ref{renyi-loss-fn-reversed} take the form,
\begin{equation}
    \partial_{\theta_k} \widetilde{D}_2(\sigma_v\|\rho) = \frac{-i{\rm Tr}\left(\left\{{\rm Tr_h}({[\widetilde{H}_k, \sigma]}),\sigma_v\right\} \rho^{-1}\right)}{{\rm Tr}\left(\sigma_v^2 \rho^{-1}\right)}. \label{gradient_sigma}
\end{equation} 
where $\widetilde{H}_k = \prod_{j=1}^{k-1} e^{-iH_j \theta_j} H_k \prod_{j=k-1}^1 e^{iH_j \theta_j}$. 
See Ref. \cite{kieferova2021quantum} for more details on the algorithms and access model needed for this calculation.
\end{document}